\documentclass{article}

\usepackage{amsmath,amssymb,amsfonts}

\usepackage{graphicx}

\usepackage{listings}

\newtheorem{definition}{Definition}  
\newtheorem{corollary}{Corollary}

\newtheorem{theorem}{Theorem}

\newtheorem{rule-of-thumb}[theorem]{Definition} 

\begin{document}
\title{Polaris: The Mathematics of Navigation and the Shape of the Earth}

\author{John P. Boyd \\
Department of Climate \& Space
Sciences and Engineering  \\University of
Michigan, 2455 Hayward Avenue, Ann Arbor MI 48109 \\
jpboyd@umich.edu}

\maketitle

\begin{abstract}
For millenia, sailors have used the empirical rule that the elevation angle of Polaris, the North Star, as measured by sextant, quadrant or astrolabe, is approximately equal to latitude.  Here, we show using elementary trigonometry that Empirical Law 1 can be converted from a heuristic to a theorem.  

 A second ancient empirical law is  that the   distance in kilometers from the observer to the North Pole, the geodesic distance measured along the spherical surface of the planet,  is the number of degrees of colatitude multiplied by 111.1 kilometers.
Can  Empirical Law 2 be similarly rendered rigorous? No; whereas as the \emph{shape} of the planet is controlled by trigonometry, the \emph{size} of our world is an accident of cosmological history. However, Empirical Law 2,  can be rigorously verified by  \emph{measurements}. The association of 111 km of north-south distance to one degree of latitude trivially yields the circumference of the globe as 40,000 km.
We also extend these ideas and the parallel ray  approximation to three different ways of modeling a 
Flat Earth. We show that photographs from orbit, taken by a very expensive satellite, are unnecessary to render the Flat Earth untenable; simple mathematics proves Earth  a sphere just as well.
\end{abstract}

  \begin{table}[h]\caption{\label{TableSymbols}  List of Symbols }
\begin{center}
{\footnotesize $\Omega$ is the angular frequncy of the Earth's rotation. When two angles add up to $\pi/2$, the angles are said to be ``complementary"; the cosine of one member of the pair is equal to the sine of the other.}

  The  ``declination [angle]" is  the angular distance of a point north or south of the celestial equator.
\begin{tabular}{|cl|} \hline
$d$ & length of a geodesic from the observer's position to the North Pole. \\  \hline
$D$ & distance from observer's position to the Pole Star, Polaris \\ \hline
$D^{star}$ & distance from observer's position to target star \\ \hline
$E^{disk}$ & equatorial circumference  \\ \hline
$E^{globe}$  & equatorial circumference  \\ \hline
$L$ & thickness of the layer of atmospheric refraction \\ \hline
$W$ & scaling for the nondimensional    distance-to-North Pole coordinate \\  \hline 
$W^{earth}$  & width (radius) of spherical earth \\ \hline
$W^{star}$ & width (radius) of star \\ \hline
$\eta$ & angle between the central ray from star to earth and the least parallel ray \\  \hline
$\epsilon$ & elevation [angle], angular  distance between the horizon and the target star. \\ & More precisely, 
 $\epsilon$ is the angle between a ray from the center of the earth \\  
     &  to the  star and a second ray of the same longitude but in the equatorial plane \\
& $\epsilon=\pi/2 - \zeta$ , synonym: ``altitude angle" \\ \hline
$\epsilon_{app}$  & apparent elevation \\ \hline
$\zeta$ & zenith  [angle],  angular  distance between directly overhead \\  & [observer's zenith] and the target star \\
  &  $\zeta=\pi/2 - \epsilon$ \\   \hline
$\varphi$ & latitude, $\varphi= \pi/2 - \theta$ \\ \hline
$\theta_{i}$ & angle of incoming ray, relative to the zenith\\
$\theta_{o}$  &  angle of outgoing ray    \\
$\theta$  & nondimensional distance-to-North-Pole    \\   \hline
$\theta$  &  colatitude, $\theta= \pi/2 - \varphi$,  [sphere only]\\  \hline
$\psi(\theta_{0}, \theta_{1})$ & angular width of arc between polar coordinates  \\
    &  $\psi(0, \theta)=\theta$ \\ \hline
$\Omega$ & $(2 \pi/86400)$ per second; angular frequency of earth's rotation. \\ \hline
  $\mathcal{M}$ & North-Pole-to-equator distance \\     \hline
$\rho$ &  air density \\ \hline
$r$ & radial coordinate in spherical coordinates   \\ \hline
$\nu$ &  Eratoshtenes parameter, $\, (W^{earth}+W^{star})/D^{star}$ \\ \hline 
$\digamma$ & refraction correction to the elevation angle \\
\hline
\end{tabular}  \end{center} \vspace{5pt} \end{table}

\section{Polaris and the True North Star}

The key measurement for navigation is the following.

\begin{definition}[elevation angle] The elevation angle $\epsilon$ is 
  the angular  distance between the horizon and the target star. More precisely, 
 $\epsilon$ is the angle between a ray from the center of the earth   
       to the  star and a second ray in the same longitude but in the equatorial plane. Alternatively, the elevation can be defined as the complementary  angle to the zenith angle $\zeta$: $\epsilon \equiv\pi/2 - \zeta$. A Synonyms for ``elevation angle include``altitude angle" and simply ``altitude".
\end{definition}

\begin{definition}[zenith angle] The zenith angle 
$\zeta$   is the  angular  distance between directly overhead  [observer's zenith] and the target star. A second
   equivalent definition   $\zeta \equiv \pi/2 - \epsilon$.
\end{definition}

The  ``Elevation angle" $\epsilon$ is ninety degrees when the Pole Star is at
zenith and zero degrees when the Pole Star is bisected by the horizon.

Both angles can be measured by a variety of ``inclinometers'' [angle-measuring devices] including the 
mariner's astrolabe, quadrant, sextant and the Arab instrument known as the \emph{Kamal}.

As explained below, angle measurements of the Pole Star are especially useful for navigation.
 However, one difficulty is that
Polaris is 3/4 of a degree from the celestial North Pole 
today, and was more than three degrees away in Elizabethan times. However, mariners soon realized 
that by looking at stars near Polaris and using the set of empirical rules with the charming name of the
``Regiment of the Stars", the distance-from-the-North-Pole errors could be greatly reduced.

\begin{definition}[Regiment of the North Star]
A set of rules that allow sightings of stars such as Kochab 
to correct for the difference between the position of Polaris and 
the true north celestial pole.
\end{definition}

We use ``Pole Star" denote a star that is not Polaris, but rather Polaris-after-the-Regiment-of-the-North-Star has been applied. Our Pole Star is thus  aligned with the
earth's rotation axis and is the true north celestial pole.

Practical star sightings incur additional errors because of the limitations of the sextant, the mariner's 
astrolable and other inclinometers. Again, correction tables and instruments were  developed many centuries
  ago \cite{Freiesleben55,Hewson51}. 
The ful twelve-minute video \cite{PracticalNavigatorPolaris}  is a careful discussion of all these corrections. For the videographer's example,
  a sextant sighting of  29 degrees, 53.5 minutes is altered to a best estimate of
 29 degrees, 21.3 minutes, a correction of about half a degree or about 30 nautical miles.

The importance of angles for navigation has driven the development of planar and 
spherical trigonometry \cite{VanBrummelen09,VanBrummelen13,VanBrummelen20,VanBrummelen21}.

\clearpage

 \section{The First Empirical Law: Latitude Is Elevation}

\begin{quote}
 For at least two millennia, navigators have known how to determine their latitude. Knowing the latitude of the desired destination, the navigator could sail north or south to that latitude and then sail east or west to reach that destination. To do this, it was necessary to have a method of measuring angles. In early days Arabs used one or two fingers at arms length to measure the angle between the horizon and Polaris. Later they used a \emph{Kamal}, which is a piece of cord with knots tied in it. This could be used to measure the angle between the horizon and Polaris. A knot could be tied in the cord as a measure of the homeport latitude before leaving, so the desired latitude was premeasured. Arabs tied knots on the cord at intervals of one \emph{issabah}, Arabic for ``finger", whic denotes 1 degree and 36 minutes.       

In the 10th century AD, Arabs introduced to Europe the astrolabe and the quadrant.
%
The quadrant spans 90 degrees and is divided into whole degrees. A plumb bob establishes the vertical. The quadrant was popular with Portuguese explorers in the 15th century. In addition to Polaris, they used observations of the Sun for determining latitude, particularly in the southern hemisphere.
\end{quote}
\hspace*{0.5in} --- P. Kenneth Seidelmann in Sec. 7 of \cite{Seidelmann1dot7}
\bigskip

Waters' treatise on navigation in the sixteenth and seventeenth centuries \cite{Waters58} discusses the Polaris-latitude connection on pgs. 41 to 50.

Royal Air Force navigators used ``latitude equals elevation" as a tool in navigating Lancaster heavy bombers in night attacks
 on Germany during World War II as described in \cite{Hoare07}, written by a retired war-time RAF navigator.

Thus, many centuries of navigation by land and sea provides overwhelming evidence for the following.

{\bf Empirical Law 1: Latitude is the Elevation Angle of Polaris}
\emph{In the Northern Hemisphere, the latitude of the observer is approximately equal to the elevation angle measured by the observer, or in other words}
\begin{eqnarray} \epsilon \approx \varphi\end{eqnarray}.

We show below that this heuristic rule is in fact a provable theorem.

Practical navigators are obliged to make small corrections to the observed or ``apparent" elevation angle to account for errors intrinsic to the sextant and aslo atmospheric refraction \cite{PracticalNavigatorPolaris}, observer height above the sea (``dip angle") and so on. The observation and corrections collectively generate  the ``true" elevation angle 
 denoted by $\epsilon$; it is this corrected elevation that appears in Empirical Law 1.

\section{The Leveled Observer}

 The surface of the planet is corrugated with small scale curvature in the form of innumerable hills and valleys. A planar earth has zero large scale curvature, but a  spherical earth is curved on a planetary scale as well as 
on a small scale. 
To obtain consistent star measurements, 
we assume a ``leveled observer" who employs a sextant with a spirit level (or equivalent surveying tricks) applied so that a vector from the surface to his zenith is parallel to the ``effective gravity" (gravity plus the centrifugal force due to the planet's rotation). 

The effective gravity is not constant; a mountain will alter the gravitational field  in its immediate vicinity. However, the variations in Earth's surface gravitational field are less than one part in 15,000 of its mean value.  The corrugations will then have no effect on our measurements or results except perhaps in the fourth decimal place. Geodesists and oil prospectors need to be concerned about these variations, but we need not.
 


\section{Empirical Law 2: The Distance to the North Pole Is the Product of Colatitude 
With 6360 Kilometers}

By the late 16th century, Englishmen knew that the ratio of distances at sea to degrees were constant along any great circle such as the equator or any meridian,
\footnote{[
A meridian is a Great Circle of constant longitude; it passes through both poles.]}
 Sssuming that Earth was a sphere. Robert Hues wrote in 1594 that the distance along a great circle was 60 miles per degree, that is, one nautical, mile per arcminute (pg. 374 of \cite{Waters58}).

 Edmund Gunter wrote in 1623 that the distance along a great circle was 20 leagues per degree. (Pg. 374 of   \cite{Waters58}), which is the same statement in a different length unit. 

Pierre-Louis Maupertuis (1698-1759) was a major figure in physics for stating the Principle of Least-Action. In 1736,
 he was appointed chief of the French Geodesic Mission to Lapland to measure the length of a degree 
of latitude. Simultneously, a second team was sent to Ecuador. The measurements triumphantly confirmed  Newton's 
prediction that Earth is a (slightly!) oblate spheroid. In 1799, the meter was defined to be  one-quarter of
the length of a meridian divided by 10,000,000.

 Let ``geodesic distance" denote the arclength of the curve that connects two points by the shortest path that lies in the surface of the planet. This definition of the meter implies that one degree of latitude shortens the 
geodesic distance to the North Pole by (10,000 kilometers/90) $\sim$  111 kilometers,  consistent with 
Hues, Gunter and other navigators. Collectively, these and other astronomers over many centuries 
arrived at the following.

{\bf Empirical Law 2: Nautical Miles to the Pole and Degrees of Latitude}
\emph{Each increase in latitude $\varphi$ by one degree decreases the geodesic distance to the North Pole from the observer by
60 nautical miles  or 
equivalently 111 kilometers.}
In symbols, this is equivalently
\begin{eqnarray}
d(\varphi) =W \, \left( \dfrac{\pi}{2} - \, \varphi \right)
\end{eqnarray}
where $W$ is a length scale determined in the next section. 
It is not possible to \emph{prove} this empirical law in the same way that Empirical Law 1 will be elevated to a theorem below. However, it is possible to prove that Emprical Law 2 can be \emph{measured} as true.

\section{A Trivial Derivation of the Radius of the Earth}

The distance $d$  along the geodesic distance (great circle arc) from the observer  to the North Pole has no obvious length scale. Nevertheless, it is still convenient to write $d$ as the product of a nondimensional distance  and an arbitrary length scale $W$.  The latitude $\varphi$ is known for millions of cities and geographical features and can be easily determined by a sextant measurement of the elevation $\epsilon(\text{Pole Star})$ via $\varphi=\epsilon(\text{Pole Star})$. It is therefore convenient to use latitude as the nondimensional 
measure of distance from the  observer to the North Pole.
The formula for $d$ becomes 
\begin{eqnarray}gin{eqnarray}~\label{vvvv}
 d(\varphi) & = &  W \, \left(  \pi/2 - \varphi \right). 
\end{eqnarray}

 Empirical Law 2 requires that  the pole-to-equator geodesic distance is the length associated with one degree 
of latitude (111.1 km) multiplied by the number of degrees of latitude (90) between the equator and 
pole, yielding $d(\pi/2)=10,000 \text{km}$. Eq.~(\ref{vvvv}) becomes
\begin{eqnarray}
10,000 \text{km}  & = & W \dfrac{\pi}{2} 
\end{eqnarray} which demands that
\begin{eqnarray}
W &= & 6360 \, \text{km} \\
  & = & W^{earth} \, 
\end{eqnarray}                                                                                                                                                                                                                                                                                                                                                                                                                                                                                                                                                                                                                                                                                                                                                                                                                                                                                                                                                                                                                                                                                                                                                                                                                                                                                                                                                                                                                                                                                                                                                                                                                                                                                                                                                                                                                                                                                                                                                                                                                                                                                                                                                                                                                                                                                                                                                                                                                                                                                                                                                                                                                                                                                                                                                                                                                                                                                                                                                                                                                                                                                                                                                                                                                                                                                                                                                                                                                                                                                                                                                                                                                                                                                                                                                                                                                                                                                                                                                                                                                                                                         
Note that $W=W^{earth}$ was obained without any presuppositions of planetary shape or curvature;
   the radius  of the spherical Earth has \emph{emerged
 spontaneously}.

\section{Parallel Ray/Distant Source  Approximation}

\begin{quote} ``The historical assumption of the sun rays parallelism
is far from being a spontaneous assumption for the students."
\end{quote} \hspace*{0.5in} --- by Nicolas D\'{e}camp and C\'{e}cile
de Hosson on pg. 919 \cite{DecampHosson12}

\bigskip

Whether light from a star falls upon the Earth or the Moon as a beam of \emph{parallel rays} or as a beam 
of \emph{divergent} rays has a profound impact upon the consequences. A perfectly parallel beam is impossible,
but a beam of \emph{almost}-parallel rays will nonetheless generate a much different situation than a beam of strongly divergent rays. It is therefore essential to quantify the meaning of ``almost-parallell".

\begin{figure}[h]
\centerline{\includegraphics[scale=0.5]{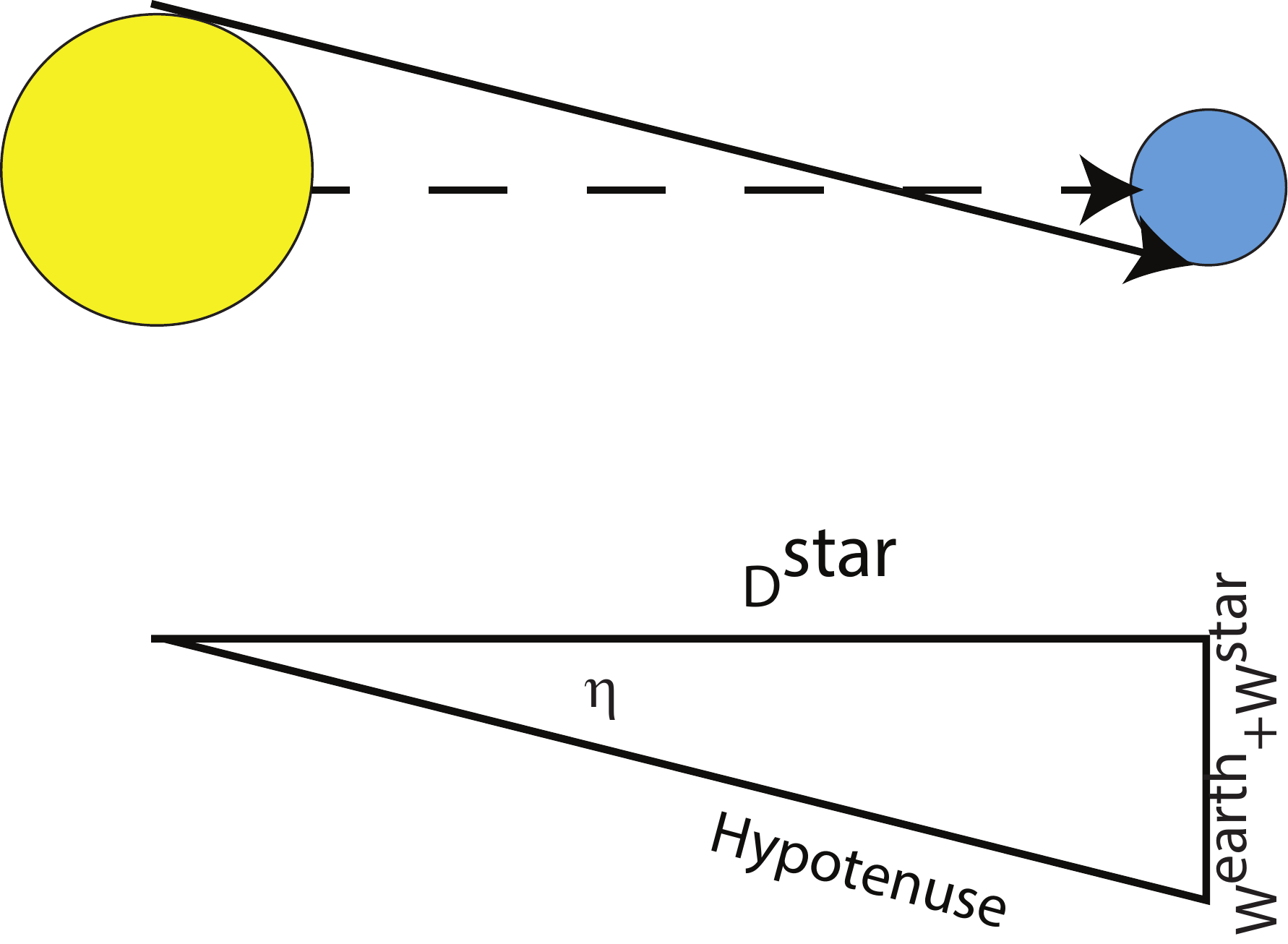}}
\caption{Top: Schematic of an emission source [star or the Sun] (yellow disk), Earth (blue disk) and light rays. The dashed line with an arrowhead is the path of a ray from the center of the star to the center of the Earth; we shall call this the ``central ray". The degree of parallelness 
of other rays is measured by the angle between the ray and the central ray. The least parallel ray that hits the Earth is one which, while traveling a distance $D^{star}$ ( horizontal in the diagram) from the star to the Earth is moving in the perpendicular direction (vertical in the figure) from one side of the star to the opposite side of the planet. This implies that this least-parallel ray is the hypotenuse of the triangle in the lower half of the diagram. The horizontal top side of the triangle is the path of the central ray with length $D^{star}$. The angle $\eta$ between these two rays is an upper bound on the angle between the central ray and all other rays that pass from star to planet.} 

\label{FigQuasiParallelRaySpheres}

 \end{figure}

Let $D^{star}$ denote the distance from the observer's position to the target star. Let $W^{star}$ and $W^{earth}$ be the widths (radii) of the light-emission source (star) and absorber (Earth).
A ray from the center of the star to the center of the surface of the earth (``central ray")  is shown as the dashed line in 
 Fig.~\ref{FigQuasiParallelRaySpheres}.

 The ray which is the ``least parallel" (of those hitting the planet) in the diagram is the solid ray with the  arrowhead. Its vertical range, as represented by the right side of the triangle in the lower half of
Fig.~\ref{FigQuasiParallelRaySpheres}, is the radius of the star plus the radius of the planet.
 Let $\eta$ denote
the angle in radians between the central ray and the least parallel ray. This is an upper bound on the angles between rays from star to planet.


\begin{definition}[Nearly Parallel Rays]
 All the rays from the star to the  Earth will be ``nearly parallel" if and only if the bound-on-angles $\eta \ll 1$.
\end{definition} 

It will prove useful to define  the ``Eratosthenes parameter" as
\begin{eqnarray}~\label{Eqnudef} 
 \nu \equiv \frac{ W^{star} + W^{earth} }  {D^{star} } \qquad 
\text{[Eratosthenes parameter]}
\end{eqnarray}

\begin{theorem}[Spread of Angles and the Eratosthenes Parameter $\nu$]
 In terms of the Eratosthenes parameter $\nu$,
\begin{eqnarray}~\label{Eqeta}
 \eta   & = & \arctan\left(  \frac{W^{star}+ W^{earth} } {D^{star} }  \right) \\
  &  =  & \arctan(\nu).  \end{eqnarray}
\end{theorem}

Proof: 
 
   The Law of Sines  applied to the triangle in Fig.~\ref{FigQuasiParallelRaySpheres} yields \cite{VanBrummelen20}
\begin{eqnarray} \eta & = & \arctan\left(\frac{W^{star}+ W^{earth} } {D^{star} } \right) ,
 \end{eqnarray} which is (\ref{Eqeta}). 
The second line follows from recognizing that the argument of the arctangent is just the
Eratosthenes parameter $\nu$ as defined by (\ref{Eqnudef}).
    $\blacksquare$

\begin{theorem}[Angles for Nearly Parallel Rays]~\label{ThAngNearlyPar}
If the distance to the star $D^{star}$ is large compared to the larger of the radius of the 
radiating star and the radius of the target planet, then

 \begin{enumerate}
\item    \begin{eqnarray} \nu \ll 1  \end{eqnarray}

\item  \begin{eqnarray} \eta \approx \nu
\end{eqnarray}

\item 
\begin{eqnarray}
\eta \ll 1
\end{eqnarray}
This implies that all rays emitted by the star and falling on the planet will be 
``nearly parallel" as defined above. 

\item In the limit $\nu \rightarrow \, 0$ or equivalently, $D^{star} \rightarrow \infty$,
the rays are parallel.
\end{enumerate}

\end{theorem}

Proof: Recall that the definition of the Eratosthenes parameter is (\ref{Eqnudef} ) $\nu \equiv (  W^{star} + W^{earth})/ D^{star}$.  When $D^{star}$ is large compared to the radii of the star and planet, it follows that $\nu \ll 1$, 
which is the first proposition.

The Taylor expansion of  the arctangent function, (4.4.42) of \cite{AbramowitzStegun65},
transforms (\ref{Eqeta}), $\eta =  \arctan\left(  \nu \right)$, into
\begin{eqnarray}
\eta & = & \sum_{j=0}^{\infty} \, (-1)^{j} \, \dfrac{1}{1+2 j} \, \nu^{1 + 2 j} \end{eqnarray}
Retaining just the term of smallest degree gives, with a relative  error of $O(\nu^{2})$,
$\eta \approx \nu$, which is the second proposition of the theorem.
 This in turn requires  that small $\nu$ implies small $\eta$, 
which is the third proposition. The fourth part of the theorem follows from taking the limit in the approximation $\eta = \nu +O(\nu^{3})$ and then recognizing that the limit of ``nearly parallel waves", differing in angle from the 
central ray by no more than $\eta$, is \emph{parallel} waves.
$\blacksquare$

  The nearly parallel approximation may be alternatively labeled the ``Distant Source" 
approximation because it becomes more and more accurate as $D^{star}$ increases.

For all  stars, $\nu$ is tiny because the distance to the star is huge compared to the diameter
of the star. 

For Polaris itself, $D^{star}= 4 \times 10^{15}$ km  and $W^{star} = 3 \times 10^{6}$ km
 yield
\begin{eqnarray}  
 \nu \approx 0.75 \,  \times 10^{-9}  \qquad 
\end{eqnarray}

For the Sun and earth combination, $D^{star}=1.496 \times 10^{8}$ km, and $W^{star}=6.96 \times 10^{5}$ km,
\begin{eqnarray}
\nu \approx 0.0046 \approx 1 / 215
\end{eqnarray}
The smallness of $\nu$ vindicates Eratosthenes, who assumed parallel (or nearly parallel) rays from the sun in his 
measurement of the radius of the earth.\cite{DecampHosson12,MrYazdanEratosthenes,Papathomas05}


\begin{figure}[h]
\centerline{\includegraphics[scale=0.25]{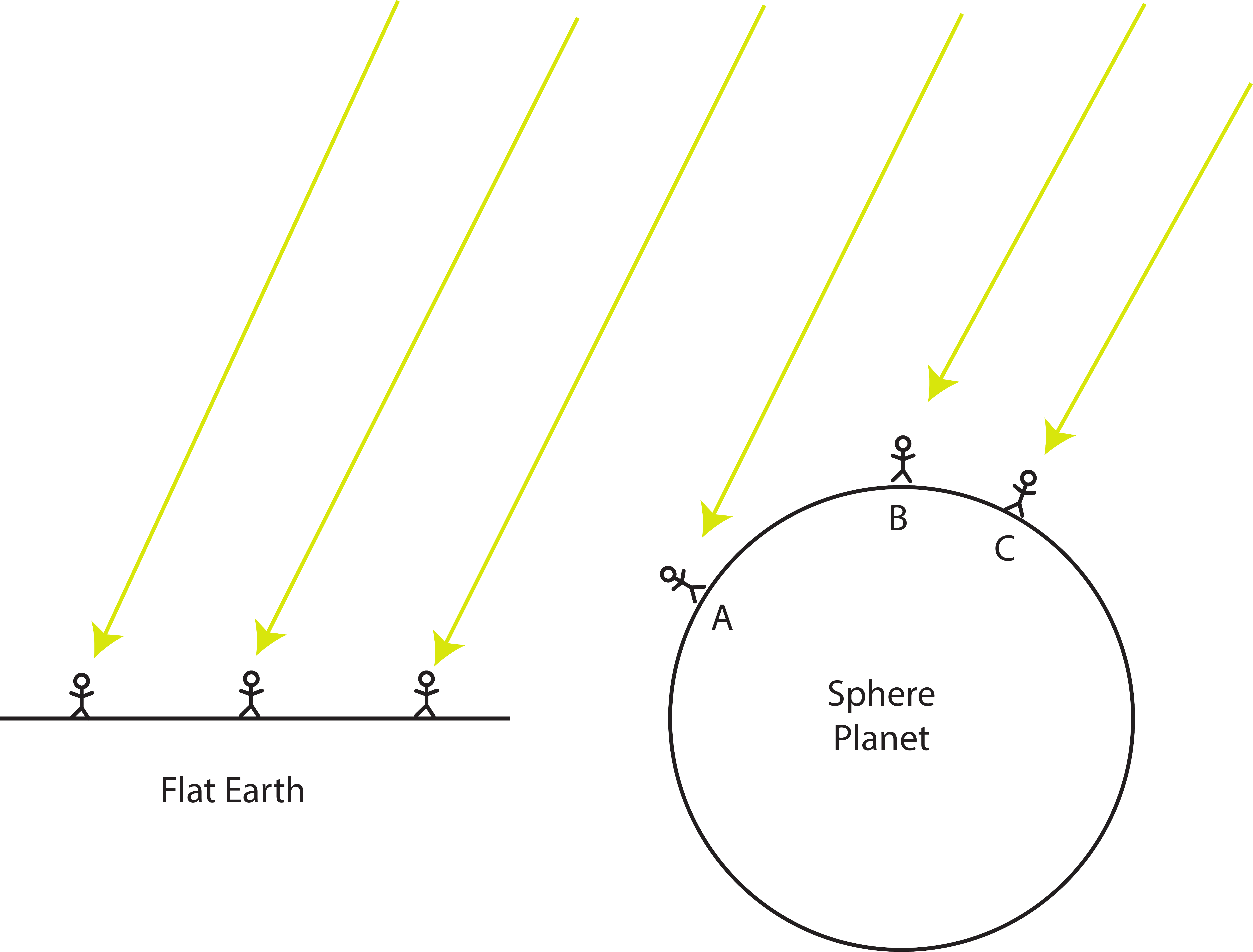}}
\caption{Parallel rays falling on a Flat Earth (left) and a spherical planet (right). All the observers
on the Flat Earth measure exactly the same elevation angle of a given star, regardless of observer location. In contrast, observers at different points on the sphere measure \emph{different} elevation angles. The elevation angle for observer C on the right is around 90 degrees where the elevation angle for observer A is near 0; for him the  rays are low, just above the horizon.} 

\label{FigPolaris_Parallel_on_Flat_and_Sphere}

 \end{figure} 

 \begin{theorem}[Flat Earth Parallel Rays] On a Flat Earth when the rays from the star are parallel, the elevation of Polaris (and every other star) is \emph{independent} of the observer's latitude.
\end{theorem}

Proof: Obvious from the left half of Fig.~\ref{FigPolaris_Parallel_on_Flat_and_Sphere}. $\blacksquare$


The Flat Earth-parallel ray  prediction that every observer sees  Polaris at 90 degrees of elevation is in gross contradiction to observations. Thus, the Flat Earth is possible only if the 
 parallel ray approximation \emph{fails},  which requires that the distance to the star is 
of comparable magnitude to  the larger  of the radii of the earth and the star.

In contrast, if the parallel ray approximation is accurate, observers on a \emph{spherical}
 earth measure \emph{different} elevation angles depending on latitude. Let us suppose that all observations are made on the same meridian as that of the target     star.

The stars appear to rotate about the Pole Star; a time-lapse photograph gives a so-called ``star trails" plot in which each star (except the Pole Star) traces a circle. Each point on the star trail gives a different angle; the desired elevation angle of a star is the angle of the point on its star trail which is closest to the observer's zenith, i. e., the highest point on the star trail.



\begin{figure}[h]
\centerline{\includegraphics[scale=0.25]{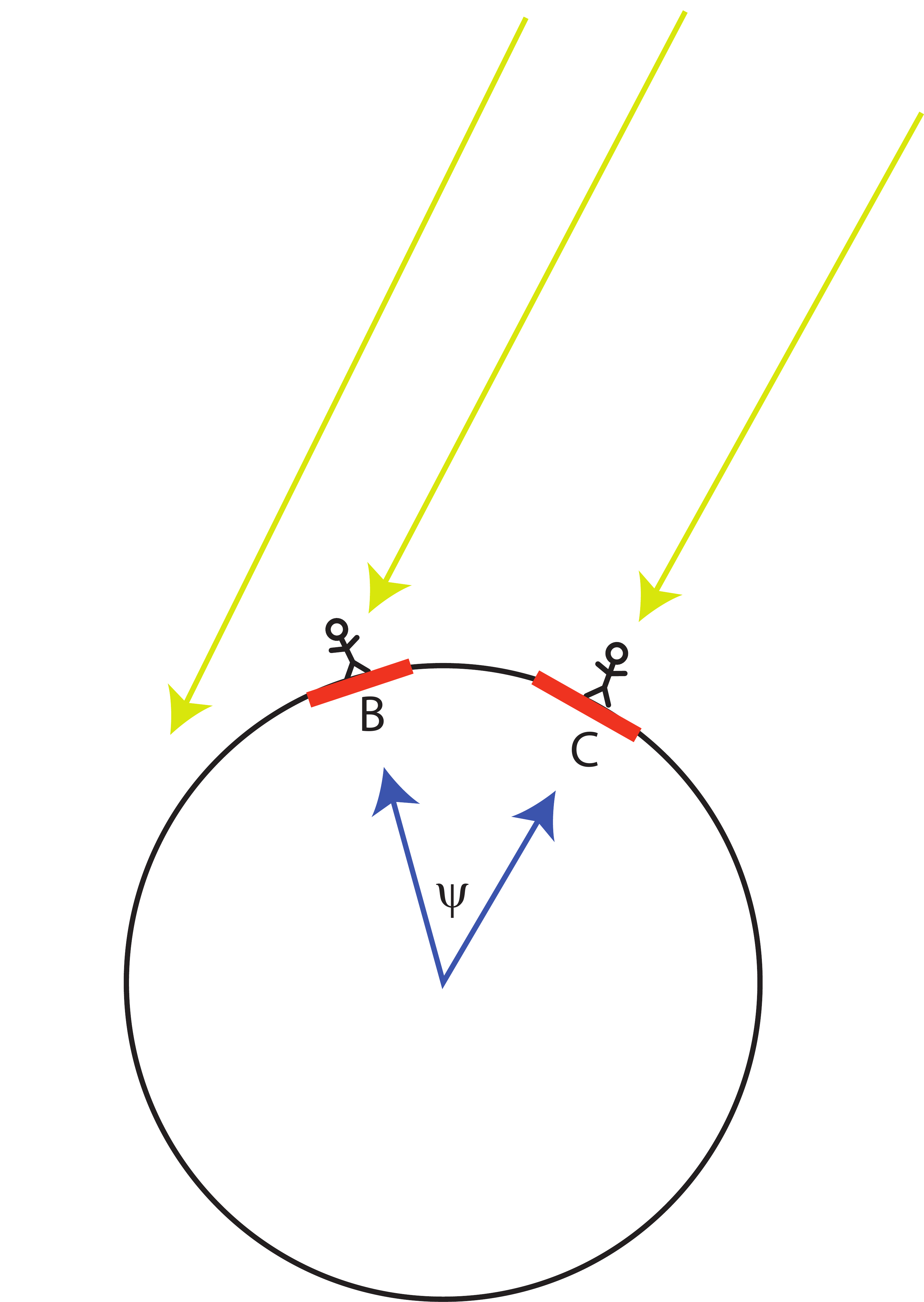}}
\caption{Parallel rays falling on a spherical planet. Two rays from the center of the planet through the observers 
define an angle $\psi$. Observers B and C will measure different elevation angles for the star. Because the rays from the star (and indeed any star) are parallel, the difference $\epsilon_{C}\, - \, \epsilon_{B}$ is due 
entirely to the fact that a line segment [red] parallel to the horizontal direction of Observer B is the same as 
for Observer C except for a rotation through the angle $\psi$.}

\label{FigPolaris_Parallel_Sphere_Observers}

 \end{figure} 

Fig.~\ref{FigPolaris_Parallel_Sphere_Observers}   shows why different observers on a meridian of a  
spherical planet obtain different angles for the beam of approximately parallel rays from a target star. The horizontal direction for each observer can be represented by a vector that is orthogonal to the 
local effective gravity vector. The crucial point is that as we move from one observer to another, this vector 
must rotate, and the rotation angle must be added to the elevation angle of the previous observer to 
obtain the elevation angle at the location of the current observer as shown in the figure. Thus, observer B measures an elevation angle $90 - \psi$ where $\psi$ is about 45 degrees while observer C has an elevation angle of
 $\epsilon_{C}= 90 \,  \text{degrees}$. The difference in elevation angle arises because B and C are separated by about one-eighth of the circumference of the planet or equivalently by an angle of 360/8 degrees. In general, the measured elevation angles of two observers differ by an angle $\Psi$ which  is 360 degrees multplied by the 
geodesic distance between the two observers divided by the circumference of the earth.


\section{Elevation Is Latitude}

\subsection{Conversion of an Empirical Law to a Theorem}

\begin{theorem}[Elevation Angle Is Latitude]~\label{ThEL1}
Let $\varphi$ denote latitude and $\epsilon$ denote the ``elevation  angle", also called 
the ``altitude angle" or ``altitude". 
If the Pole Star is observed from a spherical planet and the distance to the Pole Star is large compared to the radius of the Pole Star (allowing the parallel rays approximation), then after $\epsilon$ is corrected for the small errors of the sextant, the slight displacement of Polaris from the northern Celestial Pole, and atmospheric refraction, etc.
\begin{eqnarray}   \varphi = \epsilon \end{eqnarray}
\end{theorem}

Proof: To handle general positions of stars on the Celestial Sphere,   rotate the latitudinal coordinate so that C is at a latitude of 90 degrees N. in the new coordinate $\breve{\varphi}$. (Longitude is not modified.)    
Observer B is at a $\breve{\varphi}$ that differs from C's by $\psi$ degrees. Therefore, 
\begin{eqnarray}
\epsilon_{B} = 90 \,  \, - \, \psi \qquad    [\text{degrees}] \end{eqnarray}
But the latitude of observer B is also $90 - \psi$. It follows that latitude and elevation angle 
are always equal, which is the proposition of the theorem. 
$\blacksquare$

%
%
%
%

This theorem is identical with Empirical Law 1, which is thus converted from an empirical relationship ---
highly probable but vulnerable to disproof by as yet undisovered counterexamples  --- to a mathematical 
theorem. 

\subsection{Sextant Corrections}


As noted in the introduction, the accuracy of the sextant measurement can be    usefully improved by applying a number of 
corrections. The details can be found in any handbook of celestial navigation \cite{Cunliffe10,Burch15} or videos such as  \cite{RiggingDoctorSextant,PracticalNavigatorPolaris}; it would take us  too far afield to describe them here. Instead, the correction  for atmospheric refraction [next subsection] will have to illustrate the idea 
of corrections for all species of corrections.

The important point is that most of these errors can be greatly reduced by using the tables of the 
\emph{Astronomical Almanac}, a joint production  of the U.S. Nautical Almanac Office, U. S. Naval Observatory,   Her Majesty's Nautical Almanac Office and the U. K. Hydrographic Office. The printed version contains precise 
  positions over time of astronomical objects both natural (stars, planets, etc.) and man-made (artificial satellites) for a given year. This ephemeris  serves as a world-wide standard for such information. The online version [http://asa.hmnao.com/] extends the printed version by providing data in machine-readable form. A good source for the theory behind the \emph{Almanac} is \cite{UrbanSeidelmann12}; the history and pre-history of 
 navigational almanacs is described in \cite{SeidelmannHohenkerkWHOLE}.
The videos \cite{RiggingDoctorSextant,PracticalNavigatorPolaris} walk the reader through examples.

\subsection{Atmospheric Refraction}

When the material properties of the medium vary, this induces variations in  the speed of light, smoothly curving or discontinuously bending 
the trajectories of light rays as discussed further in Subsection~\ref{SubSecCaseIIIRefraction}.
In particular, density variations in the atmosphere alter the speed of light. 
   Unfotunately, the density of the atmosphere varies with the seasons, local weather phenomena and the exponential decrease of density required by hydrostatic equilibrium.
 Without density measurements, the best one can do is to apply a simple analytic approximation 
that represents a time-average of refraction.

Let $\epsilon_{app}$ denote the apparent elevation angle. Let $\epsilon$ denote the true  elevation (or synonymously,`` true altitude").  Let $\digamma$ denote the correction to $\epsilon_{app}$. The simplest  analytical formula for refraction is, in radians,
\begin{eqnarray}~\label{Old19}
\digamma & = &  - 0.000293     \cot(\epsilon_{app})
\end{eqnarray}
which is given on the first page of \cite{Bennett82}. It is not recommended when $\epsilon_{app}$ is
smaller than 15 degrees.

Bennett \cite{Bennett82} compares many analytic approximations for the refraction angle $\digamma(\epsilon_{app}$). We apply his method H, a more complicated formula than (\ref{Old19}), which is widely used, notably in the
 U. S.  Naval Observatory's \emph{Vector Astrometry} software.
\begin{eqnarray}
\digamma \, = 0.000291   \, \cot\left(  \epsilon_{app} + \dfrac{7.31}    {\epsilon_{app} + 4.4} \, \right)
\end{eqnarray}
Note that the argument of the cotangent function must be computed in degrees and then converted to radians.

Once the refraction $\digamma$ has been calculated, the best approximation to the elevation is
\begin{eqnarray}
\epsilon \approx \epsilon_{app} + \digamma
\end{eqnarray}
where   $\epsilon_{app}$ is the ``apparent" elevation and $\epsilon$ is the ``true'' elevation, which is the sextant measurement modified by all non-refractive corrections. This is also the best approximation to latitude, $\varphi \approx \epsilon$.

%



\begin{figure}[h]
\centerline{\includegraphics[scale=1.2]{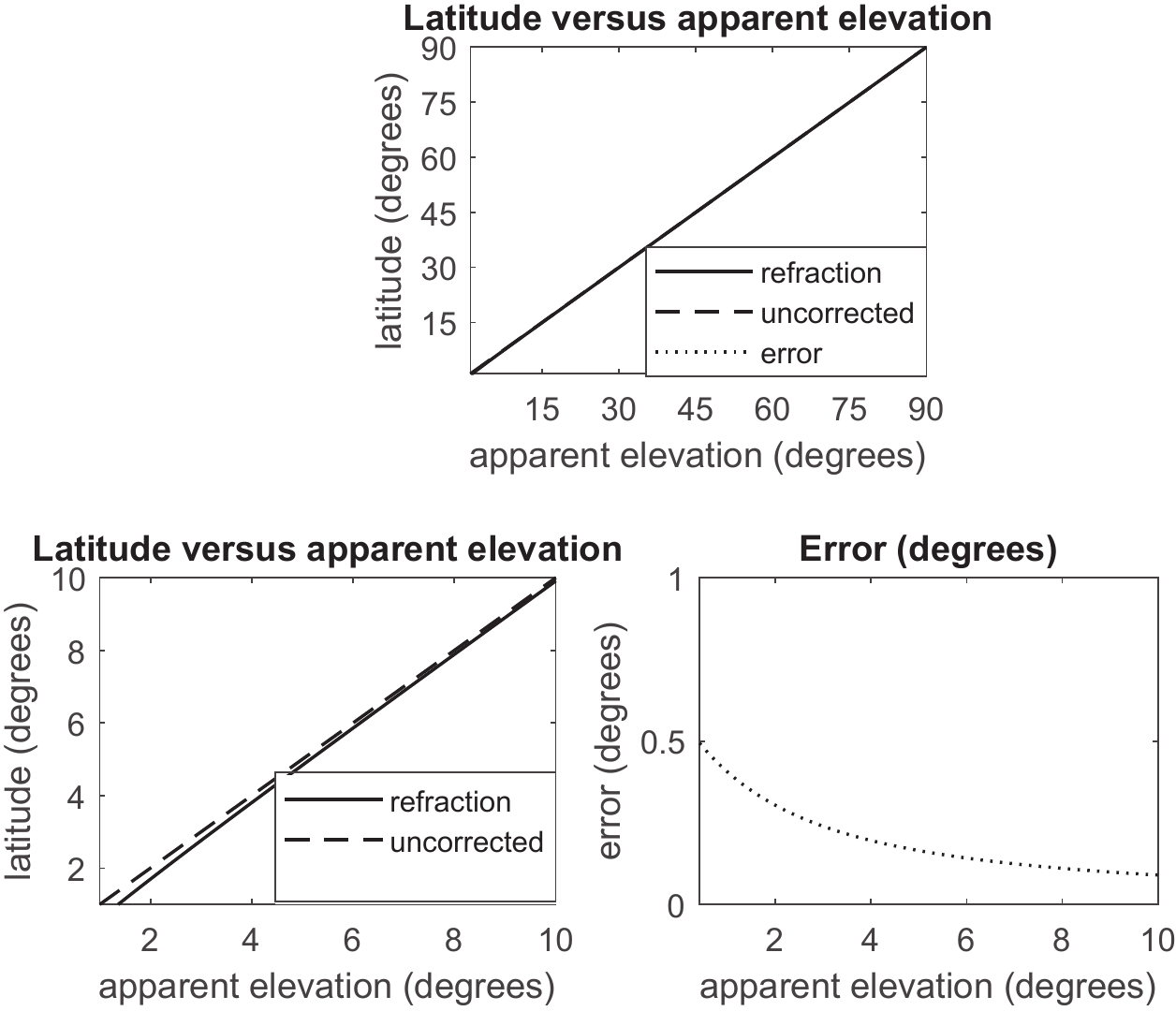}}
\caption{ Corrected and uncorrected elevation angles and their difference on the full range of the elevation angle, $\epsilon_{app } \in [0 \,\text{degrees}, 90 \, \text{degrees}]$ (top). Not that the uncorrected and refraction-corrected curves are graphically indistinguishable and the dotted error curve is invisible underneath the horizontal axis. Bottom left: the same but on the reduced range  
$\epsilon_{app } \in [0 \,\text{degrees}, 10 \, \text{degrees}]$. The lower right graph plots $ \epsilon - \epsilon_{app}$, also on the shortened  range $[0, 10 \, \text{degrees}]$.}
\label{FigElevationEqualsLatitude_RefractionB} 
\end{figure}

Fig.~\ref{FigElevationEqualsLatitude_RefractionB} and Table~\ref{OldTabSextantCorrB} show that the refractive correction is less than an arcminute
for all $\epsilon_{app}$ greater than 45 degrees, which implies a navigational error, if the refractive correction is omitted, of only one 
nautical mile. The correction grows as the elevation angle decreases, which is why navigators are advised to 
measure Polaris and other celestial objects only when the target is at least 15 degrees above the horizon.  
  However, $\epsilon$ is a good approximation to latitude even when Polaris is within a degree or two 
of the horizon. 


  \begin{table} \caption{\label{OldTabSextantCorrB} Refraction corrections using the formula ``H" of 
Bennett \cite{Bennett82}}
{\footnotesize  Note that refraction in nautical miles (rightmost column) is numerically identical to the refraction in arc minutes.}
\begin{center}
\begin{tabular}{|ccccc|} \hline
$\epsilon_{app}$ [degrees]  &  $\epsilon$ [degrees] & $\digamma$ [radians]
 & $\digamma$ [degrees] & $\digamma$ [nautical miles] 
   \\
0.5 & 0.021  &              - 0.008 &  -0.48   &  -28.8  \\
 1 &   0.59  &              -0.01     &  -0.41 &   - 24.3 \\ 
2  &  1.70 &                      -0.0071   &   -0.30  &  -18.2  \\
5 &  4.84   &                     - 0.0029 & -0.165  & -9.89    \\
10   &  9.91    &                 -  0.0016  & -0.090  &  - 5.39      \\ 
15 &  14.9  &                        -0.0011   &  -0.061 &   -3.63    \\
30  & 29.97 &                  -0.00050   & -0.029  &    -1.72 \\
45  &  44.98   &                   -0.00029  &  -0.017 &   -1.00  \\
60 & 59.99  &                 -0.00017   & -0.0096 &   -0.57   \\
75  &  75.0 &                      -0.000077   & -0.0044   &  -0.27 \\
 \hline
  \end{tabular}  \end{center} \vspace{5pt} \end{table}


\clearpage

\section{Why Flat Earth Needs a Nearby-Polaris and Non-parallel Light From the North Star} 

On a Flat Earth, the horizontal levels of all Flat Earth observers are parallel. If the light from a star like Polaris is  
also approximately parallel, then, as seen in Fig.~\ref{FigPolaris_Parallel_on_Flat_and_Sphere} \emph{all Flat  Earth observers} will see Polaris \emph{at the same point} in the celestial sphere. This 
is in gross contradistinction to what is actually observed, $\epsilon \approx \varphi$, Theorem~\ref{ThEL1}.

Therefore, Flat Earth requires that Polaris be so close to Earth --- a few thousand kilometers or less--- so that its rays are \emph{not} approximately parallel. Different observers will see 
different elevation angles.\footnote{This means jettisoning several independent measurements of the distance to Polaris by parallax and other methods that show  the actual distance to Polaris is over 400 light-years. However, a central tenet of all Flat Earth sects and denominations is that all space agencies and professional astronomers are part of a vast conspiracy and hoax, and their publications and findings must be discarded. Flat Earth must be approached as fruitful for what-if scenarios, but is otherwise the exponential of the exponential of exponential of delusion.} 
The close-Polaris hypothesis allows the computation of $D^{star}$, the distance to Polaris, by triangularization as  described in the next section.

If, on the contrary, Polaris is very distant, then the following corollary to Theorem~\ref{ThAngNearlyPar} 
shows that calculation of $D^{star}$ is impossible.

\begin{corollary}[Triangularization Fails in the Parallel Limit]
In the limit of parallel rays, it is impossible 
to obtain an estimate of $D^{star}$ from triangulation because the relevant triangles are infinitely 
long and narow, degenerating into rays. This conclusion is \emph{independent} of the shape of the planet.
\end{corollary}

Proof: Take the limit, for fixed $W^{star}$ and $W^{earth}$, of $D^{star} \rightarrow \infty$ 
in Fig.~\ref{FigQuasiParallelRaySpheres}. $\blacksquare$

\section{Triangularization On the Flat Earth for a Local Polaris}

Let $D$ denote the distance from the North Pole to the Pole Star. Equivalently, $D$ is the length of a geodesic from the star to the North Pole. Because the star is at the zenith of an observer  standing on the North Pole (a ``leveled observer" as defined earlier), the line segment is perpendicular to the surface of Flat Earth.

Let $d$ denote the distance from the observer's position to the North Pole. Let $\epsilon$ denote the elevation angle.

\begin{figure}[h]
\centerline{\includegraphics[scale=0.5]{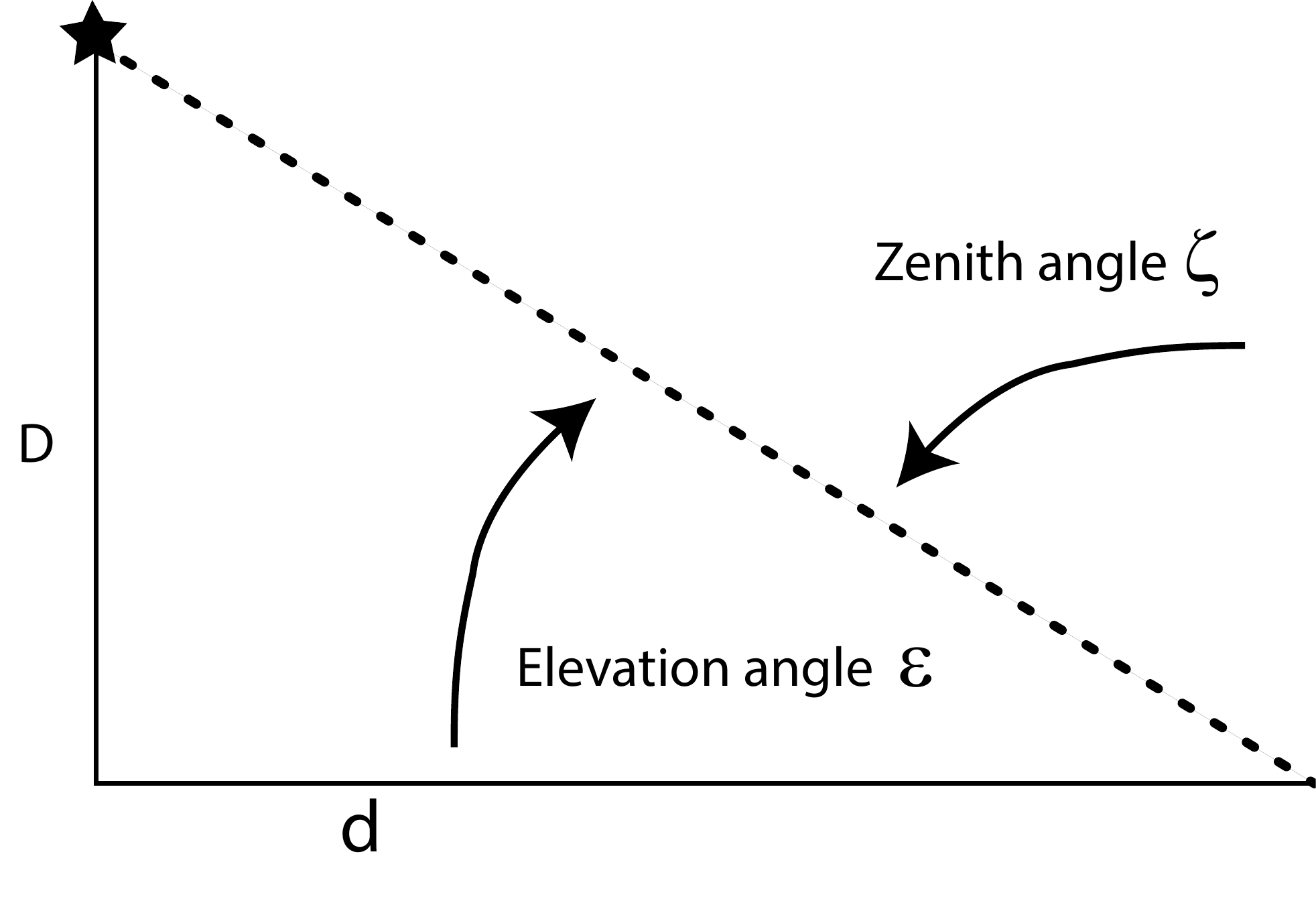}}
\caption{Triangulating the Pole Star on on a Flat Earth.} \label{FigPolaris_Triangle}

 \end{figure}

Fig.~\ref{FigPolaris_Triangle} shows a triangle whose vertices are (i) the observer's position on the surface 
of the planet (ii) the North Pole and (iii) the Pole Star. It is a good first approximation to pretend Polaris is the Pole Star  because the error is only three-quarters of a degree.


The ``Law of Triangles" \cite{VanBrummelen20} then shows that the ratio of the two shortest sides is equal to the tangent of 
the angle $\epsilon$ between these two sides, or in other words
\begin{eqnarray}~\label{EqTri}
D = d \tan(\epsilon) & \leftrightarrow &  d = D \cot(\epsilon)
\end{eqnarray}
We shall apply the left equation in Flat Earth Case I and the right equation [$d(D, \epsilon)$] in 
Flat Earth Case II.


\section{Flat Earth Case I : $d$ [distance to pole] is a linear
function of elevation angle $\epsilon$}


\begin{eqnarray}
\arctan( D/d)  \qquad  \Leftrightarrow \qquad 
 \epsilon =
\operatorname{\arctan}\left( \dfrac{ D}{d} \right)  
 \qquad \text{Pole-Star-on-the-Flat-Earth}
\end{eqnarray} 
where
$D$ is the distance to Polaris and $d$ is the distance from the
observation point to the North Pole. This can be rewritten to give the unknown $D$ in terms of the known
quantities $d$ and $\epsilon(d)$: 
\begin{eqnarray} 
D = \, d \,
\tan(\epsilon)
\qquad \Leftrightarrow   \qquad
D = d \,
\tan\left( \epsilon(d) \right)    
 \qquad \text{Pole-Star-on-the-Flat-Earth}
\end{eqnarray}

\begin{figure}[h]
\centerline{\includegraphics[scale=1.0]{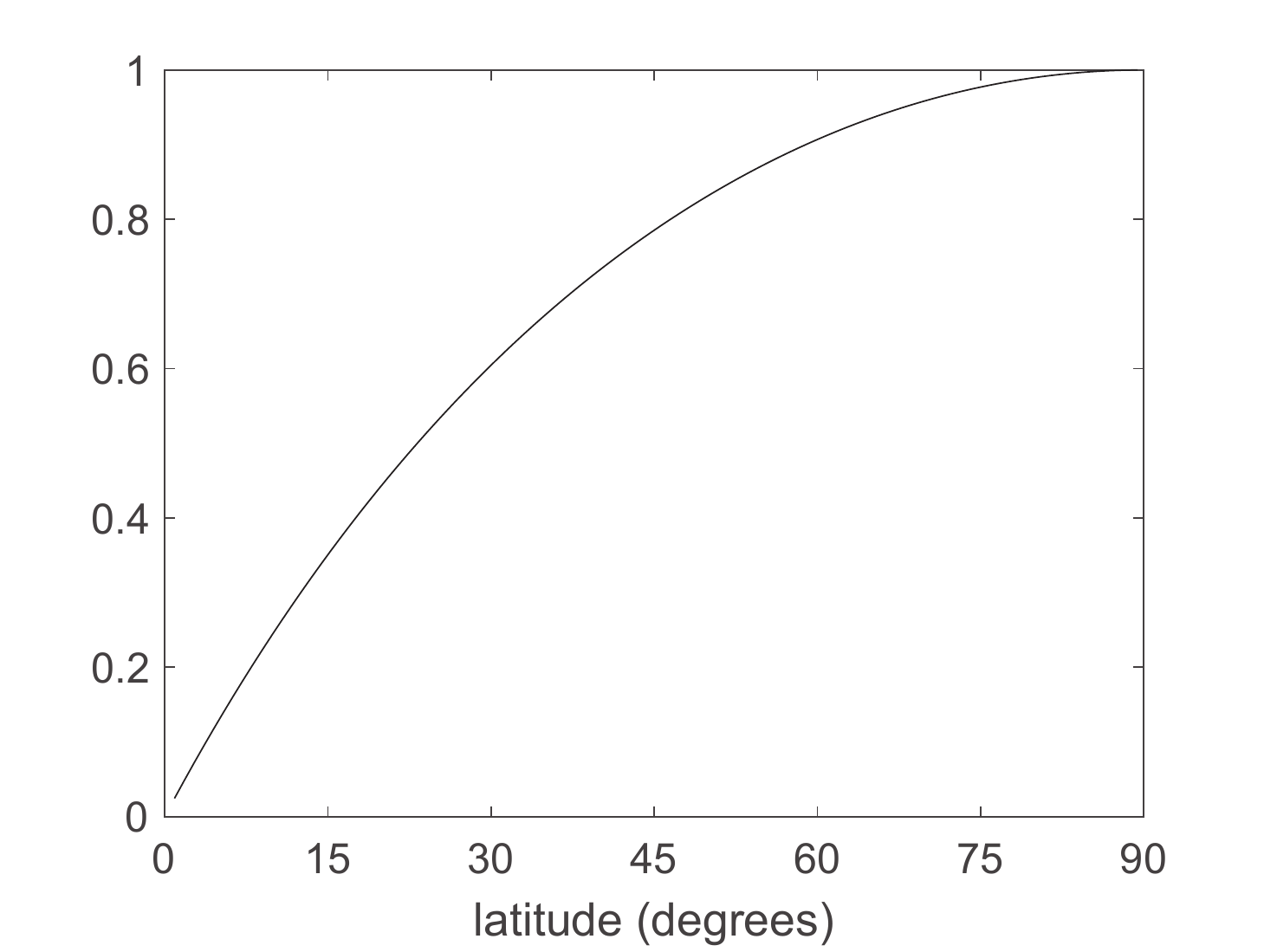}}
\caption{Flat Earth Case I: The plot shows the distance $D$ from the North Pole to the Pole
Star, scaled by $W=6360$ km, so that the nondimensional quantity plotted is $D/W= (\pi/2 - \varphi) \, \tan(\varphi)$. The figure
assumes that $d= W \, \left(\pi/2 - \varphi \right)$. }
\label{FigPolaris_FlatEarthCase_I} 

\end{figure}

Substituting 
\begin{eqnarray}
d(\varphi) & = & W \, \left(  \dfrac{\pi}{2} \, -
\, \varphi \, \right) \end{eqnarray}
 where $W= 6360$ km and replacing elevation angle $\epsilon$ by latitude $\varphi$ yields
 \begin{eqnarray}~\label{EqMoonFairy}
D(\varphi) = \dfrac{\pi}{2} W\, \left( \pi/2 -   \varphi \, \right) \,\tan\left( \varphi \, \right)
 \end{eqnarray} 
This function is plotted
in Fig.~\ref{FigPolaris_FlatEarthCase_I}.
   Table~\ref{TableFlatEarthOneOld3} lists numerical values of the distance to Polaris for different latitudes of the observer.


For a Flat Earth evangelist, the very bad news is that  Eq.~\ref{EqMoonFairy} predicts a \emph{different} 
location for Polaris for every latitude. An observer at 75 degrees N. latitude sees Polaris more than 6220 km
above the North Pole whereas an observer in the tropics at 15 degrees N. sees Polaris only 2233 km above the Flat Earth. One is reminded of the cartoon character Multiman, who could almost instantly create an army of clones.
   Even his magic powers fail here, however: placing clones of Polaris at both 2233 km and 6220 km above the planet would mean the observer now sees \emph{two} Polarises. Since the number of observers is not restricted 
to two but may be arbitrarily large, we are immediately confronted with an \emph{infinite} number of yellow giant clones of Polaris. 

To amplify the absurdity, $\max(D)=W=6360$ km. This implies that Polaris is always \emph{near} and \emph{local}, but can a star so close (with a distance of 10,000 km at most)
\begin{enumerate}
\item be a  point source
\item    be a star of the \emph{first magnitude} over the entire daylight sky
\item 
while invisible at night?
\end{enumerate}


  \begin{table}[h]
\caption{ \label{TableFlatEarthOneOld3} Distance $D$ from the North Pole to the North Star, Polaris. The star 
cannot be in multiple places at the same time, so this Flat Earth model fails.}
\vspace{5pt}
\begin{center} {\footnotesize Distances $D$ have been independently checked and confirmed in \cite{BlueMarbleScienceMitchellTrig}.}
\begin{tabular}{|c|c|c|c|} \hline  Analytical Form  &    Fourier
Latitude (degrees) $\epsilon, \varphi$ (radians)  &   $d$      &  $D$ [Distance to Polaris] \\ \hline
1 degree    & $(1/180) \pi$ & 9889     &   173 km \\
15 degrees & $\pi/12$        & 8333 km & 2233 km \\
45 degrees & $\pi/4$ &  5000 km & 5000 km \\
75 degrees & $(5/12) \pi$  & 1667 km & 6220 km \\
89 degrees &  $(89/180) \pi$ & 111.11 km & 6360 km \\  \hline
  \end{tabular}   \end{center} \vspace{5pt} \end{table}

\clearpage

\section{Flat Earth Case II: Polaris Location  $D$ is Independent of Latitude}


Polaris cannot simultaneously be at every possible distance $D$ above the North Pole from 0 to 6360 km. 
 Can this unreality be fixed?

The linear relationship between $d$, the distance from the observer to the North Pole, and latitude $\varphi$, 
which is $d(\varphi)  =  W \, \left(  \dfrac{\pi}{2} \, -
\, \varphi \, \right) $ as derived in the previous section, 
is supported  by at least half a millenia of experience at sea by those whose lives literally depended on it! Nevertheless, in the spirit that desperate illnesses often
   require drastic remedies, we now explore a  \emph{nonlinear} relationship $d(\varphi)$.

\begin{figure}[h]
\centerline{\includegraphics[scale=1.0]{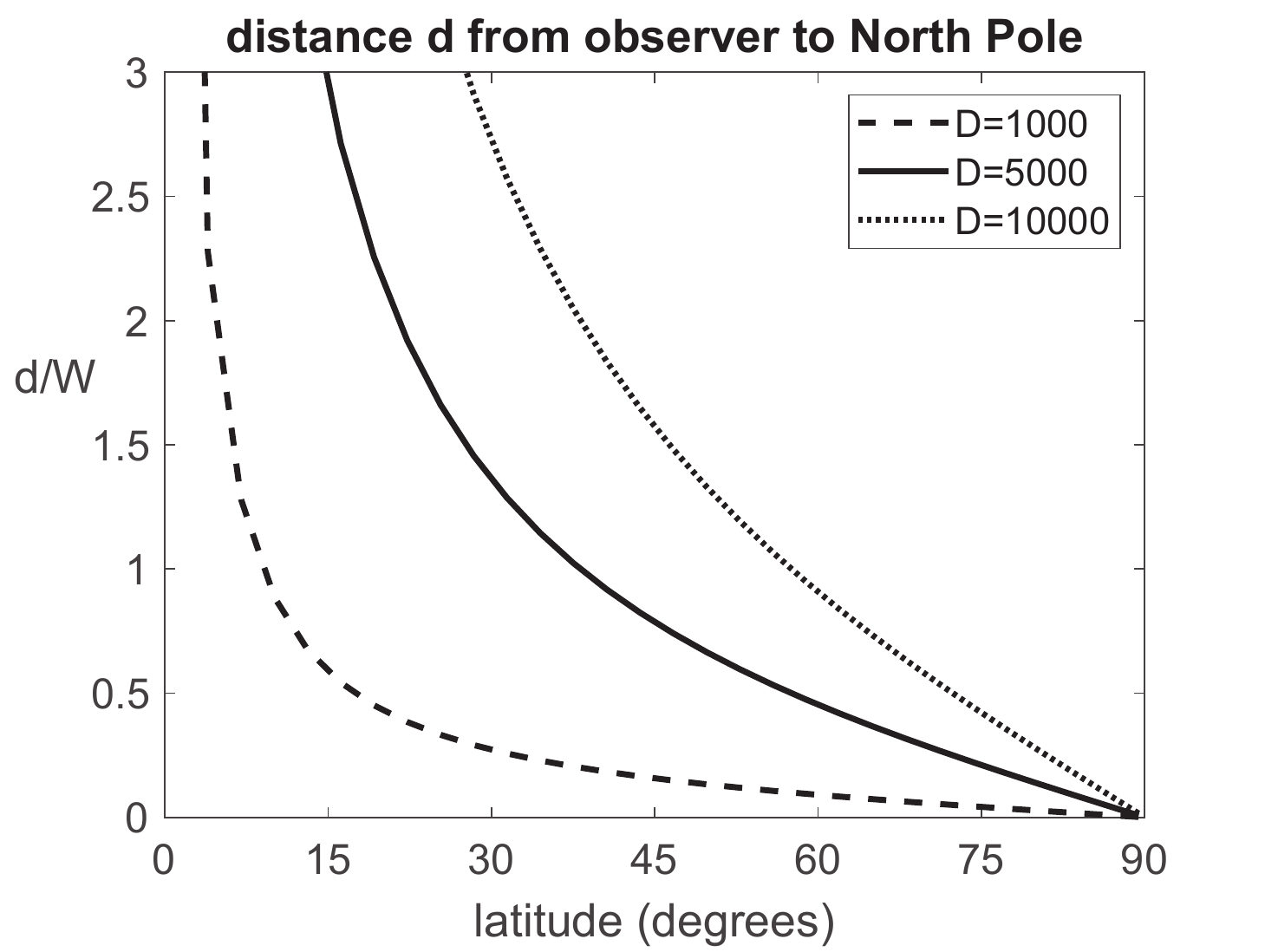}}
\caption{Flat Earth Case II.  The distance $d$ from the North Pole
to the observer at latitude $\varphi$, scaled by $W$, 
as illustrated for three different values of the distance $D$ (in kilometers) to Polaris.
}
\label{FigPolaris_FlatEarthCase_II_d_vs_varphi}


\end{figure}

The right equation in Eq.~(\ref{EqTri}) gives a relationship between $d$, $D$ and $\epsilon$. 
 If $D$ is to be a constant so that Polaris has a single, unique distance from the planet, independent of the observer's latitude, then the constraint from (\ref{EqTri}) becomes, with  $D$ as a constant,
 \begin{eqnarray} 
d(\varphi)=  D \,\cot( \varphi).
 \end{eqnarray}   as illustrated in Fig.~\ref{FigPolaris_FlatEarthCase_II_d_vs_varphi}. 
 However, as $\varphi \rightarrow 0$ (the equator), 
$d$ diverges to infinity.

 Since $d(0)=\infty$, the Northern Hemisphere, latitude $\varphi \in [0, \pi])$ is mapped to the entire infinite plane. 
 Where, then, to put the Southern Hemisphere? One could perhaps put the Southern Hemisphere on the other 
side of the infinite plane. Then a journey across the equator ---  easy in our reality --- would become a journey 
of infinite length. The Pan-American Highway from Alaska to Chile would be stretched to a never-ending strip of pavement.


We are left with insoluble difficulties:
\begin{enumerate}
\item The Northern Hemisphere is an
\emph{infinite} disk.
\item There is no place to put the Southern Hemisphere.
\item  $\lim_{\varphi \rightarrow 0} d(\varphi)= \infty$.
\item The nonlinear relationship between $d$ and $\varphi$ is refuted 
by many centuries of navigation on land, sea and air.
\end{enumerate}


\clearpage

\subsection{Flat Earth Case III: Flat Earth with Refracted Light Rays}~\label{SubSecCaseIIIRefraction}

%
%
%
%

Another possibility is that atmospheric refraction of the light from Polaris can remove the absurdities of the
two previous Flat Earth models. An important constraint is that refraction is zero in a vacuum. Therefore,
refraction will be confined to a layer of thickness $L$ where $L$ is at most a few kilometers. Equivalently, refractive light-bending occurs only in the lower and middle atmosphere. Since the distances $D$ from the North Pole to Polaris have been a few hundred to a few thousand kilometers in the earlier sections, we shall 
make the plausible assumption that 
\begin{eqnarray} L << D 
\end{eqnarray}

 We shall suppose for 
simplicity that the index of refraction (i) is equal to one, its vacuum value, everywhere above the refractive layer and (ii) varies discontinuously at $z=L$:
\begin{eqnarray}
n= \left\{ \begin{array}{c} 1,  \qquad   \, \qquad z > L \\
n_{bot}(\varphi), \qquad z \, \leq \,  L \end{array}  \right.
\end{eqnarray}
where $z$ is height above the surface of the planet.

\begin{figure}[h]
\centerline{\includegraphics[scale=0.8]{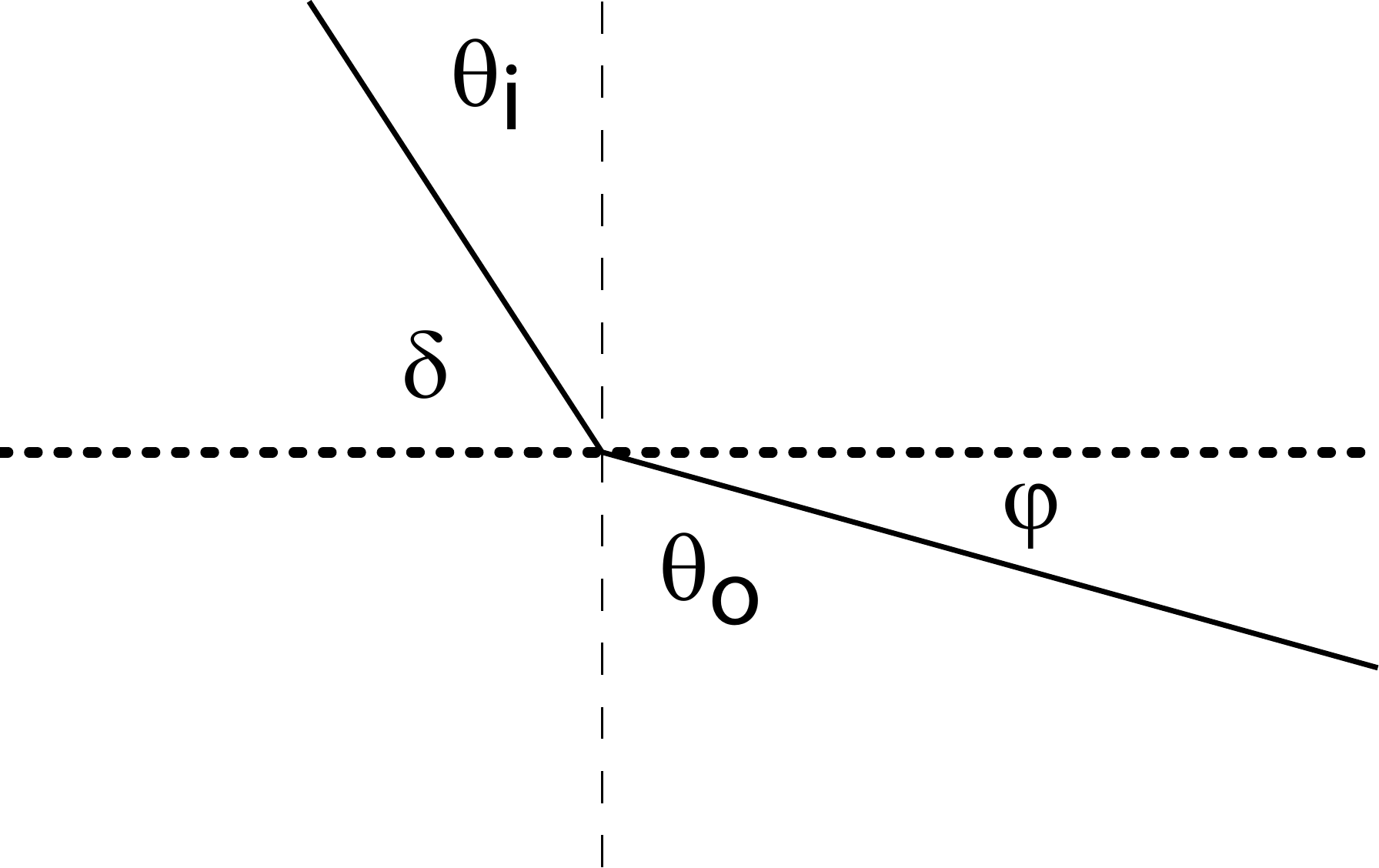}}
\caption{Snell's Law of Refraction for Flat Earth Case III for the Pole Star. The
horizontal dotted line is the boundary [interface]  between the two layers of air; the dashed line is a vertical guideline
perpendicular to the interface. $\theta_{1}$ and $\theta_{o}$  are the angles 
of the incoming and outgoing light rays. $\phi$ is colatitude           } 
\label{FigPolaris_Snells_Law}

 \end{figure}

Snell's Law is usually stated in terms of the angles $\theta_{i}$ and
$\theta_{o}$ for the ray incoming from the Pole Star and the outgoing,
refracted wave, respectively, as shown in
Fig.~\ref{FigPolaris_Snells_Law}. Denoting the indices of refraction
in the top and bottom layers by $n_{top}$ and $n_{bot}$, respectively,
Snell's law is 
\begin{eqnarray}
\dfrac{\sin(\theta_{o})}{\sin(\theta_{i})} = \dfrac{n_{top}}{n_{bot}}
\end{eqnarray}


However,  we previously employed angles
that are the complements of $\theta_{i}$ and $\theta_{o}$, $\delta=
\pi/2 - \theta_{i}$ and $\epsilon=\pi/2 - \theta_{o}$. Furthermore $n_{top}=1$. Making these substitutions and invoking the identity that the sine of an angle is equal to the cosine of its complementary angle, Snell's Law can
be rewritten  as
\begin{eqnarray}    n_{bot}(\varphi) =  \dfrac{\cos\left(\delta(\varphi)    \right)}{\cos(\epsilon(\varphi)} 
 \end{eqnarray}


To obtain  expressions for $\delta(\varphi)$ and $\epsilon(\varphi),$ we execute the following steps:
\begin{enumerate}
\item Empirical Law 1 is 
\begin{eqnarray}  \epsilon = \varphi \end{eqnarray}

\item Empirical Law 2 gives the Polaris-to-North-Pole distance as
\begin{eqnarray} d = W \, \left( \dfrac{\pi}{2} \, - \, \varphi \right)  \end{eqnarray}

\item Trigonometric identities applied to the triangle with vertices at the North Star, the North Pole 
and the observer's position gives
\begin{eqnarray}
\delta = \operatorname{arctan}\left( \dfrac{W}{ D} \left( \dfrac{\pi}{2} - \varphi) \right)\right)
\end{eqnarray}
\end{enumerate}

Then
\begin{eqnarray}  n_{bot}(\varphi) =  \dfrac{ \cos\left( 
  \operatorname{arctan}\left( \dfrac{W}{ D}        \left( \dfrac{\pi}{2} - \varphi \right)\right)
  \right)}
{\cos(\varphi)} 
 \end{eqnarray}


\begin{figure}[h]
\centerline{\includegraphics[scale=0.4]{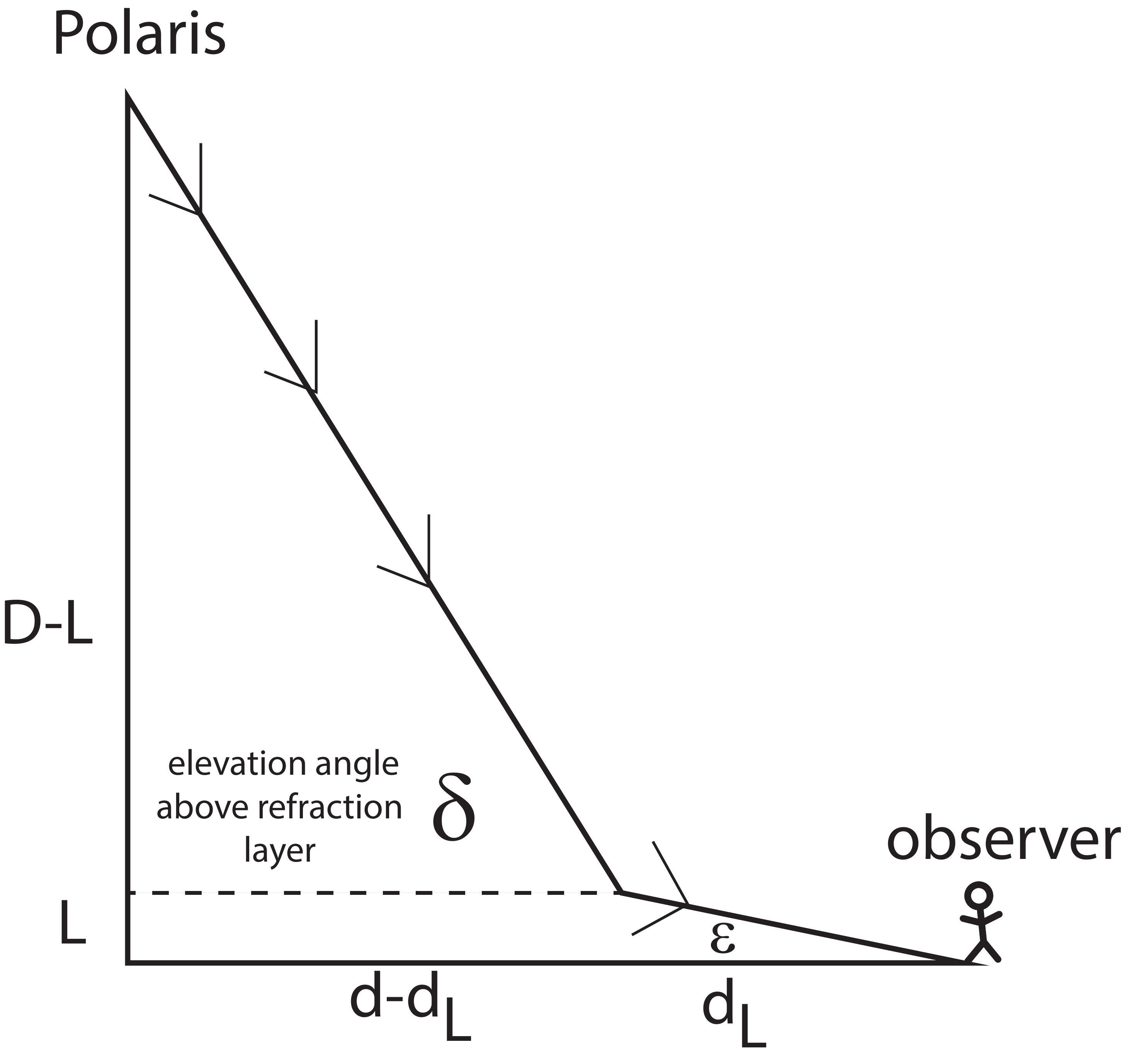}}
\caption{Flat Earth Case III. The downward-propagating light ray from Polaris (broken line segment with chevrons) has an elevation angle $\delta$ in the near-vacuum region $z \in [L, D]$. 
  At $z=L$ [dashed line], the density jumps discontinously with the density larger in the thin layer $z \in [0, L]$.   The light bends at the density jump so that the measured elevation angle is 
altered by refraction from $\delta$ (for $z > L$) to give the smaller angle
$\epsilon$ seen by an observer on the ground. The observer is a distance  $d$ from the North Pole. Note that although it is difficult to show these triangles to scale [and these triagles are therefore omitted from the graph], $D \gg L$ and $d \gg d_{L}$. These inequalities make it possible to accurately apply triangle formulas with the approximation $L=0$, i. e., the refractive layer were absent. }
\label{FigPolaris_refraction_FE_schem} 

 \end{figure}


The index of refraction observed in the atmosphere is very close to one. The
National Institutes of Science and Technology offers an Engineering
Metrology Toolbox at emtoolbox.nist.gov. This includes a calculator to
compute the index of refraction for air as a function of temperature,
humidity and pressure. One finds that even for extreme pressure, density and temperature,
the index of refraction is less than 1.004.
Furthermore, the atmosphere slows down the incoming light; the index of refraction can never be less than 1.
Thus, the physically realizable values of the index of refraction are 
\begin{eqnarray}~\label{Eqbadindex}
1  \, \leq \, n_{bot} \leq 1.004  
\end{eqnarray}

\begin{figure}[h]
\centerline{\includegraphics[scale=1.0]{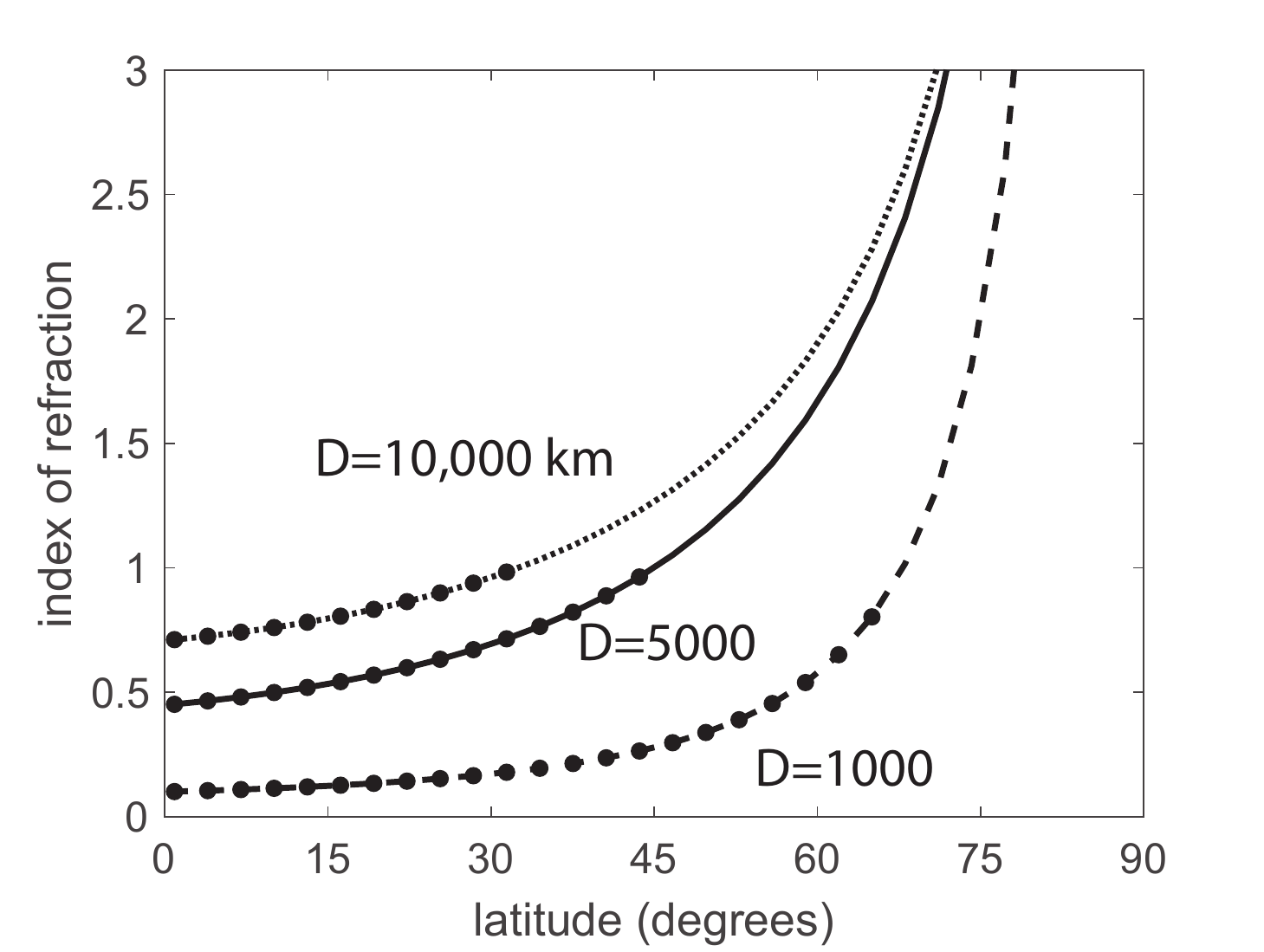}}
\caption{Flat Earth Case III. $n_{bot}(\varphi)$ for three different geodesic distances $D$ of Polaris from 
the North Pole.  Values with the index of refraction less than one, which are not physically realizable, are marked with disks.}
\label{FigPolaris_FlatEarthCase_III_Inbot_vs_varphi} 

 \end{figure}

Fig.~\ref{FigPolaris_FlatEarthCase_III_Inbot_vs_varphi} shows the required index of refraction 
in the lower, refractive layer for the Flat Earth to have a single, unique distance to Polaris. For almost all observer latitudes, the needed index of refraction $n_{bot}$ is unphysical, violating (\ref{Eqbadindex}).


A little reflection will show that the simplifying assumptions made earlier have no effect on the conclusion: Refraction fails to salvage  the Flat Earth.

    \section{Summary}

For at least half a millenia, navigators have used the empirical rule that the elevation of the Pole Star (Polaris) is equal to latitude. We have shown that proposition can be elevated to a proved theorem.

 A second ancient empirical law is  that the   distance in kilometers from the observer to the North Pole, the geodesic distance measured along the spherical surface of the planet,  is the number of degrees of colatitude multiplied by 111.1 kilometers. This cannot be proved \emph{a priori}, but we show 
that this relationship can be established by \emph{measurments}.

However, once  Empirical Law 2, that is, that distance to the pole is colatitude multiplied by 111 km, is confirmed by measurement, we prove that this rigorously 
 and uniquely determines the circumference of the Earth to be 40,000 km.

We also extend these ideas and the parallel ray  approximation to three different ways of modeling a 
Flat Earth.  All fail. As noted by many previous videographers, too numerous to list, triangularization gives different distances from Earth to Polaris for each latitude. 
  Absurd!

Photographs from space, taken on a very expensive satellite or astronauts on the moonm 
 
are unnecessary. Simple mathematics and ancient empirical laws prove Earth a sphere just as 
well.

\end{document}